\documentclass[aps,prd,nofootinbib,superscriptaddress,longbibliography,onecolumn,11pt]{revtex4-2}
\usepackage{amsmath, amssymb, mathrsfs, amsfonts, graphicx}
\usepackage{hyperref}
\usepackage{orcidlink}

\makeatletter
\AtBeginDocument{\immediate\write\@auxout{\string\citation{apsrev42Control}}}
\makeatother

\begin{document}

\title{Vacuum Spin-Torsion Energy Density and Two-Modulus Stability on
$S^2\times\mathbb{C}P^2$}

\author{Edward J. Shaya\ \orcidlink{0000-0002-3234-8699}}
\email{eshaya2@gmail.com}
\affiliation{Department of Astronomy, University of Maryland,
College Park, MD 20742, USA}
\email{eshaya2@gmail.com}

\begin{abstract}
In Einstein--Cartan--Kaluza--Klein compactification on
$S^2\times\mathbb{C}P^2$, elimination of the non-propagating torsion
generates a quartic spin-current contact interaction whose vacuum
expectation value is nonzero even at zero fermion occupation.  We
reduce the ten-dimensional $\Gamma_{ABC}$ structure to the zero-mode
sector and find that the reduction is channel-resolved: each of the
three internal Cartan-plane bilinears contribute with the opposite
sign to the external (axial) channel, so that the full trace is
$R_{\rm ch} = T_{\rm vac}/T_{\rm axial} = 1 + 3\,(-1) = -2$.  The
vacuum term is therefore twice the axial estimate and opposite in
sign, and, combined with the elimination constant
$c_{10} = -\kappa_{10}/32$, verified by explicit Clifford
computation, it is attractive and proportional to $N_{\rm fam}^2$.
We obtain the separated-point correlator, which in the continuum
scales as $N_{\rm fam}^2 d_R/(V_6\,r^6)$, and, conditional on the
physical quantum-boundary prescription, evaluate it at the minimum
supported separation.  The finite-density piece of the same correlator
has the same sign and grows with chemical potential, so binding
deepens after nucleation; the single-particle Hartree self-contraction
vanishes identically for chiral zero modes.  
Using the derived $1/(a_1^2a_2^4)$ vacuum term together with the
Einstein-frame curvature and quantized-flux contributions, we then
analyze a coupled two-modulus toy potential.  At fiducial couplings, the
$(m,n)=(3,3)$ sector has a finite-radius stationary point with positive
Hessian in both radion directions, while the four-family $(4,3)$
minimum lies beyond the adopted quantum-curvature boundary.  The
alternative three-family sector $(1,5)$ is also locally stable but is
less deeply bound in the same truncation.  Three-family dominance of
the thermal nucleation weight further imposes the quantitative
condition $\kappa_4 \lesssim 7.5\,(16/B)\,\ell_P^2$.  The stability
result is explicitly a curvature--flux--torsion toy-model result rather
than a proof against arbitrary quantum corrections.  These results serve as
inputs to the higher-level synthesis in Ref.~\cite{ShayaUECKK2026}.
\end{abstract}

\maketitle

\section{Introduction}
\label{sec:intro}

In the Einstein--Cartan formulation of gravity coupled to spinors,
torsion is algebraically determined by the spin current and can be
eliminated, generating a four-fermion contact interaction
\cite{Kibble1961,HehlDatta1971}.  On
$X_6 = S^2\times\mathbb{C}P^2$, the chiral zero modes constructed in
the companion paper~\cite{ShayaZeroModes2026} carry maximal internal
spin bilinears in every chirality plane, fixed by the same alignment
that makes them zero modes, and the contact operator inherits a rich
channel structure from the ten-dimensional Clifford algebra.  The
immediate question is, therefore, not merely whether torsion is present,
but whether the zero-mode sector turns that algebraic interaction into
a definite vacuum contribution with a calculable dependence on the
family number.

This question is the dynamical bridge between the two companion
calculations and the UECKK synthesis of Ref.~\cite{ShayaUECKK2026}.
The zero-mode analysis supplies a map from the flux integers $(m,n)$ to
the exact chiral multiplicity
$N_{\rm fam}=m(n^2-1)/8$.  The present paper asks how that multiplicity
feeds back on the compactification dynamics.  We show that the
connected spin-current fluctuations generate an attractive term
proportional to $N_{\rm fam}^2/V_6$, determine its channel coefficient,
and follow the same operator after the zero modes become occupied.
When this derived term is combined with internal curvature and
quantized flux, the family number is no longer only a spectral label:
it becomes a parameter of the two-modulus potential.  Ref.~\cite{ShayaUECKK2026}
uses precisely this link to compare the disjoint flux sectors of
$S^2\times\mathbb{C}P^2$ and to ask which sectors remain both stabilized
and semiclassically admissible.  In that higher-level account, the
$N_{\rm fam}^2$ torsion pull deepens the wells as the family number
grows, while the unequal flux support of the $S^2$ and
$\mathbb{C}P^2$ factors drives the latter toward the quantum-curvature
boundary.  Thus, the significance of this class of manifolds is not
only that their Dirac index can reproduce three families, but that the
same topological multiplicity enters a calculable gravitational vacuum
term and thereby participates in the dynamics that distinguishes the
flux sectors.

Two results are emphasized at the outset because they are
prescription-independent.  First, the separated-point correlator of
the contact density scales in the continuum as
$N_{\rm fam}^2\,d_R/(V_6\,r^6)$; this is unconditional.  Second, the
channel structure of the reduction---the factor
$R_{\rm ch} = -2$ relative to the axial channel---is fixed by the
Clifford algebra and the zero-mode spin texture alone.  Only the
\emph{evaluation} of the correlator at a definite short-distance
scale, and hence the numerical coefficient of the vacuum energy
density, is conditional on a physical short-distance prescription; we
adopt the quantum-boundary ($\mathcal{B}_Q$) prescription of
Ref.~\cite{ShayaUECKK2026} for that step and keep the two layers
strictly separated.  The later stability and nucleation calculations
are therefore applications of a derived field-theoretic input rather than
part of the index theorem itself.

Conventions: $\ell_P = 1$; metric signature $(-,+,\ldots,+)$; Dirac
conjugate $\bar\Psi = \Psi^\dagger A$ with
$A\,\Gamma^M A^{-1} = -(\Gamma^M)^\dagger$; internal normalizations
follow Ref.~\cite{ShayaZeroModes2026}
($\hat R_{S^2}=2$, $\hat R_{\mathbb{C}P^2}=24$,
$V_6 = 4\pi a_1^2\,V_{\mathbb{C}P^2}$).  All index sums over
antisymmetric multi-indices are \emph{unrestricted} (each ordered
triple counted $3!$ times); ratios of channel traces are
normalization-independent and are quoted as such.

\section{Torsion elimination: the Lorentzian interaction Hamiltonian}
\label{sec:elimination}

The first physical question is whether the spin--torsion interaction is
attractive or repulsive in the Lorentzian Hamiltonian.  In
Einstein--Cartan gravity, the torsion equation is algebraic rather than
propagating, so eliminating the connection yields a local nonlinear
spinor interaction; this mechanism dates back to the gauge-theoretic
formulation of gravity and the Einstein--Cartan--Dirac equation
\cite{Kibble1961,HehlDatta1971}.  The sign of the resulting
Hamiltonian density decides whether torsion acts as an additional
source of gravitational attraction on the internal space or as a
repulsive core resisting compactification; since every later
conclusion inherits this sign, we fix it first, before any point
splitting, vacuum evaluation, or Euclidean continuation, where sign
conventions are easiest to audit.

The gravitational Lagrangian with totally antisymmetric contorsion
$K_{MAB}$ and a Dirac spinor is
\begin{equation}
\mathcal{L} = \frac{1}{2\kappa_{10}}\,R[\omega+K]
+ \frac{i}{2}\big(\bar\Psi\Gamma^M D_M\Psi - \overline{D_M\Psi}\,
\Gamma^M\Psi\big).
\label{eq:L10}
\end{equation}
The contorsion enters quadratically through the curvature and linearly
through the spin connection.  Explicit evaluation (verified by direct
matrix computation on the $32\times32$ Clifford representation) gives
\begin{align}
R[\omega+K] &= R[\omega] + \nabla(\cdots)
- K_{MAB}K^{MAB},
\label{eq:QK}\\
\mathcal{L}_{\rm ferm} &\supset K_{MAB}\,J^{MAB},
\qquad
J^{MAB} = \tfrac{i}{4}\,\bar\Psi\,\Gamma^{MAB}\,\Psi,
\label{eq:coupling}
\end{align}
using $\{\Gamma^M,\Gamma^{AB}\} = 2\,\Gamma^{MAB}$; the bilinear
$J^{MAB}$ is real in the stated conjugation convention.  Eliminating
$K$ ($K_{MAB} = \kappa_{10}J_{MAB}$) yields
\begin{align}
\mathcal{L}_{\rm contact}
&= \frac{\kappa_{10}}{2}\,J_{MAB}J^{MAB}
 = -\,\frac{\kappa_{10}}{32}\,\mathcal{O},
\label{eq:L-contact}\\
\mathcal{O}
&\equiv
(\bar\Psi\Gamma_{ABC}\Psi)
(\bar\Psi\Gamma^{ABC}\Psi).
\label{eq:O-def}
\end{align}
with unrestricted index sums.  Since $\mathcal{O}$ contains no
derivatives, the interaction Hamiltonian density follows before any
Euclidean continuation:
\begin{equation}
\mathcal{H}_{\rm int} = -\,\mathcal{L}_{\rm contact}
= +\,\frac{\kappa_{10}}{32}\,\mathcal{O},
\qquad
\rho_{\rm EC} = \frac{\kappa_{10}}{32}\,
\langle\mathcal{O}\rangle .
\label{eq:H-int}
\end{equation}
The sign of the energy density is thus the sign of the Lorentzian
expectation value $\langle\mathcal{O}\rangle$, evaluated below with
spacelike point-splitting; Euclidean continuation is used only as a
computational device and is stated as such where it appears.

\section{Channel structure of the zero-mode reduction}
\label{sec:channels}

Eliminating torsion gives a ten-dimensional four-fermion operator, but
that operator does not yet tell us how the compactification zero modes
feel it.  As with any contact interaction, the
Nambu--Jona-Lasinio model is the familiar example; the physics
depends on which channels the operator feeds after decomposition, and
different channels can enter with different signs.  Here, the
decomposition is dictated by the ten-dimensional Clifford algebra: the
reduction distributes the interaction between a purely
four-dimensional axial channel and mixed spacetime--internal channels,
weighted by internal spin matrix elements of the zero modes.  
We therefore determine the channel content on the actual
zero-mode subspace before computing any vacuum expectation value; this determines
whether the topologically selected chiral modes amplify or
cancel the familiar axial Einstein--Cartan interaction.

Decomposing $\Psi(x,y) = \sum_I\psi_I(x)\otimes\eta_I(y)$ over the
zero modes, the $\binom{10}{3}$ antisymmetric triples split by
internal index count.  Triples with one or three internal indices
vanish on the zero-mode sector due to internal chirality.  The surviving
channels are:

\emph{External (axial) channel} ($A,B,C$ all spacetime):
$\Gamma^{abc}$ is dual to $\Gamma_d\Gamma_5$, and the channel reduces
to the square of the four-dimensional axial current.

\emph{Mixed channels} ($A$ spacetime; $B,C$ internal):
$\Gamma^{a\,ij} = \gamma^a\otimes\lambda^{ij}$, a four-dimensional
vector current times an internal spin bilinear.  On the zero-mode
subspace, only the three Cartan-plane pairs survive, because the
internal spinor structure of the zero mode is fixed by its very
existence.  Writing the six internal directions in three chirality
planes ($S^2$ and the two K\"ahler planes of $\mathbb{C}P^2$) with
generators $\Sigma_{(k)}=-i\lambda_a\lambda_b|_{(k)}$, the zero-mode
spinor $e_+\otimes e_{00}$ satisfies
\begin{equation}
\Sigma_{(1)}=\Sigma_{(2)}=\Sigma_{(3)}=+1\,.
\label{eq:internal-spin}
\end{equation}
All odd internal bilinears therefore vanish by chirality, while the
three two-index (Cartan-plane) bilinears are maximal.  This
alignment---the same alignment that makes the mode a zero mode---means
the zero-mode field subspace carries \emph{maximal internal spin
polarization in each chirality plane, independently of whether the
modes are occupied}.

The relative sign of the two channels can be seen without any
computation.  For a Weyl fermion, the axial and vector currents
coincide up to sign, so both channels reduce to the same
four-dimensional current correlator and differ only by their internal
factors.  The external channel's internal factor is
$\operatorname{tr}[\Pi_0^2] = +1$.  Each mixed Cartan-plane channel
carries instead the \emph{square of an internal expectation value of an
anti-Hermitian generator}: with
$\Sigma_{(k)} = -i\lambda_a\lambda_b|_{(k)}$ Hermitian and equal to
$+1$ on the zero mode,
\begin{equation}
\langle\lambda_a\lambda_b\rangle_{(k)} = +\,i
\;\Longrightarrow\;
\langle\lambda^{ab}\rangle\langle\lambda_{ab}\rangle_{(k)}
= (i)(i) = -1 .
\label{eq:sign-flip}
\end{equation}
Each Cartan plane therefore contributes the \emph{same} spacetime
trace as the external channel but with opposite sign, and there are
three planes:
\begin{equation}
\boxed{\;
R_{\rm ch} \equiv \frac{T_{\rm vac}}{T_{\rm axial}}
= 1 + 3\times(-1) = -2 \; }
\label{eq:Rch}
\end{equation}
The pattern
$T_{\rm axial} : T_{\rm mixed} : T_{\rm vac} = +4 : -12 : -8$ found by
explicit matrix enumeration of all triples (in both Euclidean and
Lorentzian signature, with identical result) is exactly this
structure; the Clifford computation serves as the check, not the
derivation.  The ratio $R_{\rm ch} = -2$ is independent of the
multi-index normalization convention.

Two consequences follow.  The vacuum term is \emph{twice} the
axial-channel estimate, and since the internal spin texture is a
property of the zero-mode subspace itself, the mixed channels are
active in the vacuum; no occupation is required.

\section{The separated-point correlator (continuum result)}
\label{sec:correlator}

The next question is whether an unoccupied chiral sector can contribute
to binding at all.  The mean spin current of the vacuum vanishes,
$\langle J\rangle = 0$, but its fluctuation does not,
$\langle J\!\cdot\!J\rangle_{\!c} \neq 0$, and it is the fluctuation
that the quartic torsion operator measures.  The situation is
Casimir-like: just as an unoccupied electromagnetic field exerts
measurable forces through its zero-point fluctuations, the unoccupied
zero-mode sector contributes a vacuum energy through the connected
two-point function of its composite current.
Separated-point products are also the cleanest way to keep the universal
short-distance behavior distinct from the prescription used at
coincidence; point splitting has a long history as a regulator of local
quantum operators and vacuum polarization \cite{Schwinger1951}.  We
therefore compute the current correlator at finite spacelike separation
first and postpone the physical ultraviolet prescription to the next
section.

The Wick contraction of $\mathcal{O}$ over the zero-mode sector
factorizes into the internal overlap contractions of
Ref.~\cite{ShayaZeroModes2026}
($\sum\mathcal{G}_{IJJI} = N_{\rm fam}^2/V_6$), the channel factor
$R_{\rm ch}$, and the four-dimensional current correlator.  For a free
Weyl multiplet of dimension $d_R$, the Euclidean two-point function is
\begin{equation}
\langle j_\mu(x) j_\nu(0)\rangle_E
= -\,\frac{d_R}{2\pi^4 r^6}\,I_{\mu\nu},
\qquad
I_{\mu\nu} = \delta_{\mu\nu} - \frac{2\,r_\mu r_\nu}{r^2},
\label{eq:jj-euclid}
\end{equation}
with $\delta^{\mu\nu}I_{\mu\nu} = 2$.  Continued to Lorentzian
signature at \emph{spacelike} separation, the physical
point-splitting for the Hamiltonian density
Eq.~\eqref{eq:H-int}, the $\eta$-contracted correlator is
\begin{equation}
\eta^{\mu\nu}\langle j_\mu(x) j_\nu(0)\rangle
= -\,\frac{d_R}{\pi^4\,r^6},
\qquad r \;\text{spacelike},
\label{eq:jj-lorentz}
\end{equation}
negative-definite.  Assembling with the external-channel duality
factor ($\mathcal{O}_{\rm ext} = -6\,j_5\!\cdot\!j_5$ for
unrestricted sums) and Eq.~\eqref{eq:Rch},
\begin{equation}
\big\langle \mathcal{O}(x,x')\big\rangle_{\!c}
= -\,\frac{12\,d_R\,N_{\rm fam}^2}{\pi^4\,V_6\;r^6}\,,
\label{eq:O-separated}
\end{equation}
valid for $M_{\rm KK}^{-1} \lesssim r \ll L_4$.  This separated-point
result is the unconditional content of the paper: it involves no
short-distance prescription.  With Eq.~\eqref{eq:H-int}, the vacuum
contribution is \emph{attractive} at every supported separation.

It is important to distinguish this connected separated-point
correlator from a conventionally normal-ordered local operator.
Normal ordering at exact coincidence can, by definition, subtract the
free-vacuum contraction.  The claim made here is different: the
physical short-distance prescription of Sec.~\ref{sec:bq-eval}
defines the composite interaction by a finite supported separation (or
an equivalent cell average), so the connected vacuum fluctuation is
not discarded by a coincidence subtraction.  Accordingly, the sign
and $r^{-6}$ scaling of Eq.~\eqref{eq:O-separated} are
prescription-independent, whereas interpreting its finite-cell value
as a vacuum energy density is conditional on the quantum-boundary
prescription.

\section{Quantum-boundary evaluation (conditional)}
\label{sec:bq-eval}

A separated correlator fixes the sign and scaling of the interaction, but
a vacuum energy density requires a rule for how closely the two operator
insertions may be brought together.  Ordinary effective field theory
answers this by renormalization and matching; minimum-length arguments
in quantum gravity instead question whether arbitrarily short proper
distances are physical at all \cite{Mead1964}.  The Quantum Boundary
Framework of Ref.~\cite{ShayaUECKK2026} supplies such a physical endpoint.
This section, therefore, makes the one explicitly conditional step in the
calculation: evaluating the otherwise continuum result at the minimum
supported geometric separation.

To convert Eq.~\eqref{eq:O-separated} into an energy density requires
a short-distance prescription.  In the continuum, the coincidence
limit is defined by renormalization and matching at
$\mu_{\rm match}\sim M_{\rm KK}$, leaving a scheme-dependent constant.
The quantum-boundary framework replaces this with a physical
statement: separations below $\ell_P$ are not supported by the
configuration space, and the operator is evaluated at the minimum
supported separation,
\begin{equation}
\rho_{\rm EC}
= \frac{\kappa_{10}}{32}\,
\big\langle\mathcal{O}\big\rangle_{\!c}\Big|_{r=\ell_P}
= -\,\frac{3}{8}\,C_{\mathcal{B}}\,d_R\;
\frac{\kappa_{10}}{V_6\,\ell_P^6}\;N_{\rm fam}^2 ,
\label{eq:rho-BQ}
\end{equation}
where $C_{\mathcal{B}}$ encodes the smearing prescription and equals
$1/\pi^4$ for the hard point-split at $r=\varpi\ell_P$, with $\varpi$
the boundary sharpness parameter of Ref.~\cite{ShayaUECKK2026}; the coefficient
$3/8 = 12/32$ contains the single factor of the correlator
normalization, the channel factor, duality factor, and elimination
constant are separately accounted and no factor of $\pi^{-4}$ enters
twice.  With $d_R = 16$,
\begin{equation}
\lambda_{\rm vac}
\equiv \frac{|\rho_{\rm EC}|}{N_{\rm fam}^2}
= \frac{6}{\pi^4}\;\frac{\kappa_{10}}{V_6\,\ell_P^6}
= \frac{6}{\pi^4}\,\kappa_4(V_6)\;\ell_P^{-6},
\label{eq:lam-vac}
\end{equation}
where $\kappa_4(V_6) = \kappa_{10}/V_6$ is \emph{moduli-dependent}:
the fundamental constant is $\kappa_{10}$, and quoting a
four-dimensional $\kappa_4$ requires specifying the internal volume at
which it is referred.  The $1/V_6$ scaling of
Eq.~\eqref{eq:rho-BQ} reproduces exactly the
$1/(a_1^2 a_2^4)$ moduli dependence of the torsion term in
Ref.~\cite{ShayaUECKK2026}, now with a derived coefficient.  The
evaluation point $r=\varpi\ell_P$ introduces an explicit factor
$\varpi^{-6}$ into $\lambda_{\rm vac}$; within the admissibility window
$\varpi\in(0.95,1.04)$ established in Ref.~\cite{ShayaUECKK2026} this
varies $\lambda_{\rm vac}$ by $\pm30\%$, well inside the
$\mathcal{O}(1)$ prescription uncertainty, so it does not affect the
viability condition Eq.~\eqref{eq:viability}.

Two systematic caveats bound the conditional result.  First,
$C_{\mathcal{B}}$ is prescription-dependent at the $\mathcal{O}(1)$
level (hard split versus smeared kernel).  Second, at
$r = \ell_P$ the separation is comparable to $M_{\rm KK}^{-1}$, and
the restriction of the internal propagator to the zero-mode sector is
not parametrically justified: the massive Kaluza--Klein tower
contributes at relative order $e^{-M_{\rm KK}r}\sim\mathcal{O}(1)$.
Flux backgrounds can shift degeneracies and eigenvalues of the
nonzero modes.  The leading high-frequency part of the tower
contribution is expected to be more local and less sensitive to the
index than the exact zero-mode contribution, so substantial
cancellation in the sector \emph{differences} that control the
nucleation weights is plausible; the flux-dependent part of the
Kaluza--Klein tower nevertheless remains an $\mathcal{O}(1)$
uncertainty on $\lambda_{\rm vac}$, which we flag accordingly.

\section{Finite-density correlator}
\label{sec:finite-mu}

Vacuum binding is only useful cosmologically if populating the chiral
zero modes does not reverse their signs once matter appears.  Dense fermion
systems modify propagators through occupation of the Fermi sea, and the
separation of vacuum and density pieces is standard in finite-density
field theory \cite{FreedmanMcLerran1977}.  We therefore repeat the current
contraction at nonzero chemical potential and ask a deliberately narrow
question: does Pauli blocking, the removal of occupied states from
the vacuum contraction, weaken, cancel, or deepen the spin--torsion
attraction at the post-nucleation scale?

\begin{figure*}[t]
\centering
\includegraphics[width=\linewidth]{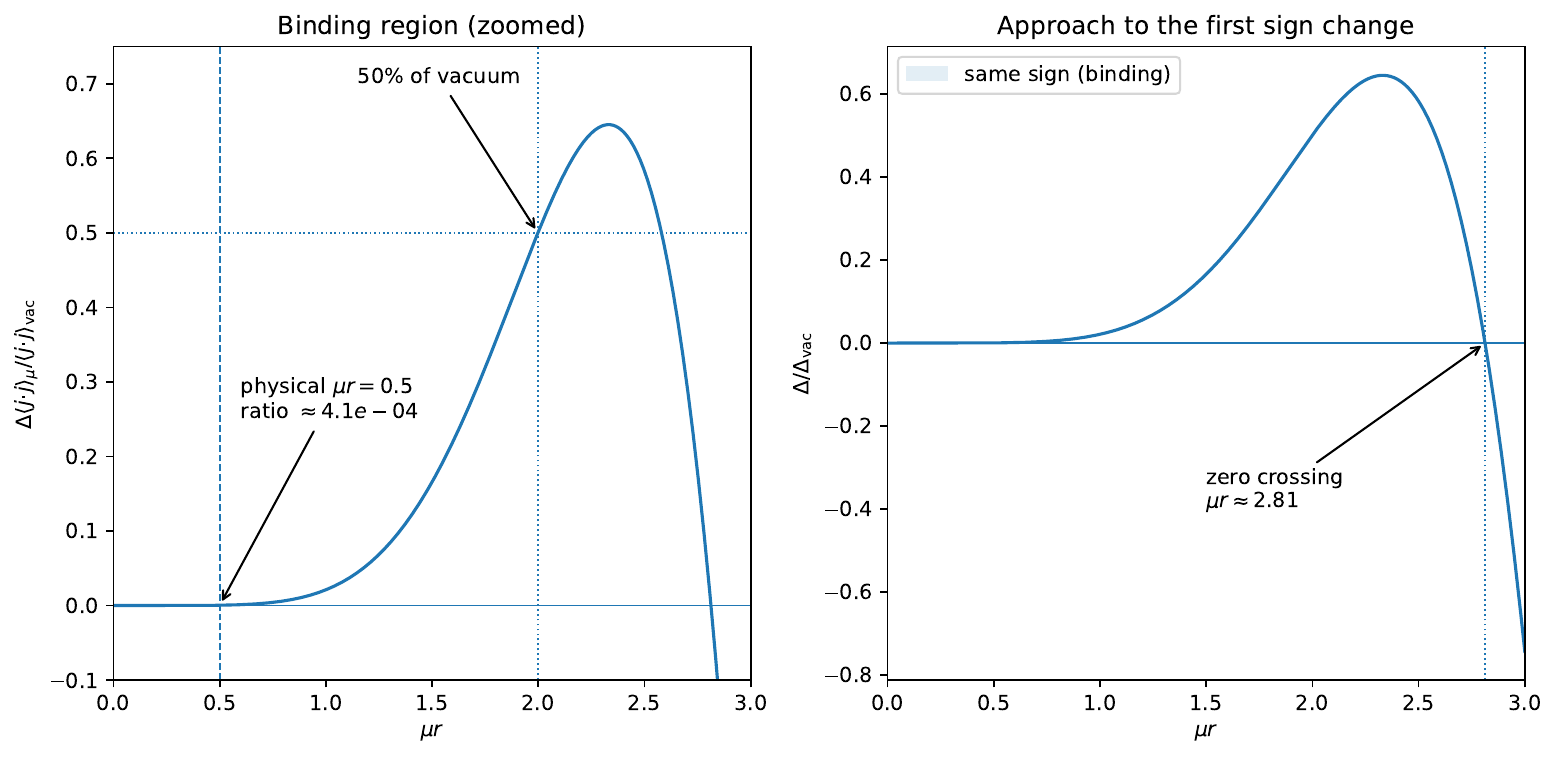}
\caption{Finite-density correction to the connected chiral-current
correlator, normalized by the zero-density vacuum contribution, as a
function of the single dimensionless variable $\mu r$.  Positive values
have the same sign as the attractive vacuum term and therefore deepen the
binding.  Both panels show the same interval, $0\leq\mu r\leq3$: the left
panel magnifies the small positive correction, while the right panel shows
the approach to the first zero crossing at $\mu r\simeq2.813$.  The
fiducial marker $\mu r=0.5$ follows from the representative occupation
scale $\mu\simeq M_{\rm KK}\simeq0.5\,\ell_P^{-1}$ together with the
minimum supported point-splitting distance $r\simeq\ell_P$ used in the
Quantum Boundary Framework.  At that point, the finite-density correction
is $4.06\times10^{-4}$ of the vacuum term; it reaches $50\%$ of the vacuum
magnitude at $\mu r\simeq1.998$.  All numerical markers are derived from
Eq.~\eqref{eq:finite-mu-ratio}, not fitted.  The source archive includes
\texttt{plot\_finite\_mu\_correction.py}, with the occupation scale and
supported splitting radius exposed as command-line arguments.}
\label{fig:finite-mu}
\end{figure*}

After nucleation, the cascade populates the zero modes, and the
correlator acquires a finite-density piece.  For free massless Weyl
fermions at zero temperature and chemical potential $\mu$ (particles
only; the post-baryogenesis universe carries net fermion number), the
equal-time Wightman functions decompose over the radial integrals
\begin{align}
I_0(\mu,r) &= \frac{1}{2\pi^2}\int_0^\mu\! dp\,p^2 j_0(pr)
= \frac{\sin\mu r - \mu r\cos\mu r}{2\pi^2 r^3},
\nonumber\\
I_1(\mu,r) &= \frac{1}{2\pi^2}\int_0^\mu\! dp\,p^2 j_1(pr)
= \frac{2 - 2\cos\mu r - \mu r\sin\mu r}{2\pi^2 r^3},
\label{eq:I0I1}
\end{align}
where $j_{0,1}$ are spherical Bessel functions, $r$ is the spatial
point-splitting, and $\mu r$ is the single dimensionless control
parameter.  The two contraction blocks are
$W = \tfrac12(I_0 - i\,\sigma\!\cdot\!\hat r\,I_1)$ and
$V = \tfrac12(I_0 + i\,\sigma\!\cdot\!\hat r\,I_1)$ (plus the vacuum
antiparticle block).  Carrying out the same $\eta$-contracted Weyl-current
trace as in Sec.~\ref{sec:correlator} gives the normalized correction
\begin{align}
\frac{\Delta\langle j\!\cdot\!j\rangle_\mu}
     {\langle j\!\cdot\!j\rangle_{\rm vac}}
&=\frac14\Big\{[\sin x-x\cos x]^2 \\
&\qquad-[2-2\cos x-x\sin x]^2\Big\},
\qquad x\equiv\mu r .
\label{eq:finite-mu-ratio}
\end{align}
All factors of $d_R$, $N_{\rm fam}$, $V_6$, and $r^{-6}$ cancel in
this ratio.  Figure~\ref{fig:finite-mu} plots Eq.~\eqref{eq:finite-mu-ratio}.
The fiducial point $\mu r=0.5$ is singled out because the compactification
analysis uses a representative occupation scale $\mu\simeq M_{\rm KK}
\simeq0.5\,\ell_P^{-1}$, while the Quantum Boundary Framework evaluates
the regulated correlator at the minimum supported separation
$r\simeq\ell_P$.  Their product therefore yields $\mu r\simeq0.5$.
At this point, the finite-density correction is only
$4.06\times10^{-4}$ of the vacuum contribution.  More generally, the
correction remains same-signed for $\mu r\lesssim2.813$ and reaches half
the vacuum magnitude at $\mu r\simeq1.998$.  Thus, the dominant binding is
already present in the vacuum term; finite occupation produces only a small
same-signed deepening at the fiducial post-nucleation scale.

Two conclusions.  (i)~The matter piece is same-signed: binding is
present at nucleation and \emph{deepens} with occupation; no sign flip
between epochs occurs or is needed.  (ii)~The single-particle Hartree self-contraction vanishes
identically for massless chiral zero modes: for a single mode
$j_5 = \pm j$ is lightlike, so the Lorentz-contracted square of its
own bilinear $\bar u\,\Gamma^{ABC} u$ is null (verified for all
channels).  This does not by itself exclude a collective
polarized-spin contribution from the ensemble; it removes only the
single-particle self-energy piece, and the binding is carried by the
connected exchange correlator computed above.  The collective
contribution, if present, is precisely the finite-density correction
$\Delta\langle j\!\cdot\!j\rangle_\mu$: it is
same-signed, subdominant at the physical occupation
$\mu\sim M_{\rm KK}\sim 0.5\,\ell_P^{-1}$ ($\Delta/\Delta_{\rm vac}\approx4\times10^{-4}$, sub-percent),
and already included in the analysis.  The
matter-era binding is therefore itself a quantum-fluctuation
effect, the finite-density part of the same connected
correlator, and classical spin-fluid (Weyssenhoff) modeling is
inapplicable to chiral zero modes.

\section{Coupled two-modulus stability test}
\label{sec:stability}

An attractive vacuum term is not by itself a compactification mechanism:
a monotonic attraction can simply collapse the internal space.  Flux
compactification provides the classic countervailing effect; quantized
field strength grows rapidly as a compact factor shrinks and can balance
curvature \cite{FreundRubin1980,RandjbarDaemi1983}.  In Kaluza--Klein language, the two-factor radii are radions, and the
physically relevant test is two-dimensional in this radion space:
after adding the derived torsion term to curvature and flux, does the
coupled $S^2\times\mathbb{C}P^2$ potential possess a finite-radius
stationary point, and are \emph{both} radions stabilized there? Is
the Hessian positive in every direction, not merely along the
isotropic ray?

The vacuum correlator fixes the moduli dependence of the torsion term:
using $V_6\propto a_1^2a_2^4$, Eq.~\eqref{eq:rho-BQ} contributes
$-C N_{\rm fam}^2/(a_1^2a_2^4)$ to the four-dimensional
Einstein-frame potential.  We now ask the limited but concrete
question whether this attraction can coexist with a finite-radius
minimum once the standard internal-curvature and quantized-flux terms
are retained.  This is a stability test within a specified
curvature--flux--torsion truncation, not a proof of all-orders quantum
stability.

For the product metric
\begin{equation}
 ds_{10}^2=e^{2\alpha}\tilde g_{\mu\nu}dx^\mu dx^\nu
 +a_1^2\hat g_{S^2}+a_2^2\hat g_{\mathbb{C}P^2},
\end{equation}
with $\hat R_{S^2}=2$, $\hat R_{\mathbb{C}P^2}=24$, and the
Einstein-frame Weyl factor $e^{2\alpha}\propto(a_1^2a_2^4)^{-1}$,
the minimal two-modulus potential has the form
\begin{equation}
\begin{split}
U(a_1,a_2)=&-A\left[\frac{2}{a_1^4a_2^4}
 +\frac{24}{a_1^2a_2^6}\right]\\
&+B\left[\frac{m^2}{2a_1^6a_2^4}
 +\frac{q n^2}{a_1^2a_2^8}\right]
-C\frac{N_{\rm fam}^2}{a_1^2a_2^4},
\end{split}
\label{eq:U2mod-spin}
\end{equation}
where $A,B,C>0$, $q=\mathcal O(1)$, and
\begin{equation}
N_{\rm fam}=m\frac{n^2-1}{8}.
\end{equation}
The three terms are respectively the Einstein-frame internal-curvature
attraction, quantized flux pressure, and the spin--torsion vacuum
binding derived above.  Isotropically they scale as $d^{-8}$,
$+d^{-10}$, and $-d^{-6}$.  Thus, the attractive terms drive
contraction at large volume, while the steeper positive flux term
provides the short-distance support against collapse.

To display the minimization explicitly, define logarithmic radion
coordinates $x_i=\ln a_i$.  A monomial
$T_\alpha=c_\alpha a_1^{-p_\alpha}a_2^{-q_\alpha}$ then satisfies
\begin{equation}
 \partial_{x_1}T_\alpha=-p_\alpha T_\alpha,
 \qquad
 \partial_{x_2}T_\alpha=-q_\alpha T_\alpha.
 \label{eq:monomial-gradient}
\end{equation}
The two stationarity conditions obtained from
Eq.~\eqref{eq:U2mod-spin} are therefore
\begin{align}
0={}&\frac{8A}{a_1^4a_2^4}
 +\frac{48A}{a_1^2a_2^6}
 -\frac{3Bm^2}{a_1^6a_2^4}
 \nonumber\\
&-\frac{2Bqn^2}{a_1^2a_2^8}
 +\frac{2CN_{\rm fam}^2}{a_1^2a_2^4},
\label{eq:stationary-x1}\\
0={}&\frac{8A}{a_1^4a_2^4}
 +\frac{144A}{a_1^2a_2^6}
 -\frac{2Bm^2}{a_1^6a_2^4}
 \nonumber\\
&-\frac{8Bqn^2}{a_1^2a_2^8}
 +\frac{4CN_{\rm fam}^2}{a_1^2a_2^4}.
\label{eq:stationary-x2}
\end{align}
These equations show directly how the two flux sectors enter the two
radion forces with different powers and coefficients.  They are solved
simultaneously for positive $(a_1,a_2)$ in each discrete $(m,n)$
sector.

The logarithmic-radion Hessian is
\begin{equation}
 H_{ij}=\frac{\partial^2U}{\partial x_i\partial x_j}.
 \label{eq:log-hessian}
\end{equation}
For each monomial above, its contribution is especially simple:
\begin{equation}
 H^{(\alpha)}=
 T_\alpha
 \begin{pmatrix}
 p_\alpha^2 & p_\alpha q_\alpha\\
 p_\alpha q_\alpha & q_\alpha^2
 \end{pmatrix}.
 \label{eq:monomial-hessian}
\end{equation}
At a stationary point, $H$ and the Hessian in $(a_1,a_2)$ are related
by the congruence transformation
$H=J^T H_a J$, with $J=\mathrm{diag}(a_1,a_2)$.  Hence, positive
definiteness is coordinate independent.

As an explicit example, for $(m,n)=(3,3)$ and the fiducial Planck-unit
benchmark $A=C=1$, $B=16$, and $q=1$, solving
Eqs.~\eqref{eq:stationary-x1} and \eqref{eq:stationary-x2} gives
\begin{equation}
 (a_1^*,a_2^*)=(2.2995,\,2.0924)\,\ell_P.
 \label{eq:33-radii}
\end{equation}
At that point,
\begin{equation}
 H_{(3,3)}\simeq
 \begin{pmatrix}
 0.5800 & 0.3767\\
 0.3767 & 1.7231
 \end{pmatrix},
 \label{eq:33-hessian}
\end{equation}
whose eigenvalues are $0.4670$ and $1.8361$.  Thus, the stationary
point is a local minimum in both radion directions.  Repeating the same
procedure in the neighboring flux sectors gives
Table~\ref{tab:spin-stability}.

\begin{table}[t]
\caption{Local minima of the curvature--flux--torsion toy potential
$U(a_1,a_2)$ in Eq.~\eqref{eq:U2mod-spin}, evaluated at the fiducial
Planck-unit couplings $A=C=1$, $B=16$, and $q=1$.  The flux integers
$(m,n)$ determine the chiral family count
$N_{\rm fam}=m(n^2-1)/8$.  The columns give the stabilized radii
$a_1^*$ and $a_2^*$, the resulting internal Kretschmann scalar
$\mathcal K_{\rm int}=4/a_1^4+192/a_2^4$, and the two eigenvalues of the
logarithmic-radion Hessian $H_{ij}=\partial^2U/\partial x_i\partial x_j$
with $x_i=\ln a_i$.  Positive eigenvalues in both directions establish
local stability within this two-modulus truncation.}
\label{tab:spin-stability}
\begin{ruledtabular}
\begin{tabular}{cccccc}
$(m,n)$ & $N_{\rm fam}$ & $a_1^*$ & $a_2^*$ &
$\mathcal K_{\rm int}$ & $\operatorname{eig}(H)$\\
\hline
$(1,3)$ & 1 & 1.700 & 2.646 & 4.40 & $(0.105,0.391)$\\
$(2,3)$ & 2 & 2.151 & 2.349 & 6.50 & $(0.189,0.745)$\\
$\mathbf{(3,3)}$ & $\mathbf{3}$ & $\mathbf{2.299}$ & $\mathbf{2.092}$ & $\mathbf{10.16}$ & $\mathbf{(0.467,1.836)}$\\
$(4,3)$ & 4 & 2.366 & 1.896 & 14.99 & $(1.046,4.090)$\\
$(1,5)$ & 3 & 1.300 & 2.873 & 4.22 & $(0.400,1.431)$\\
\end{tabular}
\end{ruledtabular}
\end{table}

Here
\begin{equation}
\mathcal K_{\rm int}=\frac{4}{a_1^4}+\frac{192}{a_2^4}.
\end{equation}
The positive Hessian eigenvalues establish genuine local minima within
the truncation.  In particular, the $(3,3)$ three-family sector is not
merely an index-theoretic configuration: the same spin--torsion term
derived from the zero-mode vacuum is compatible with a finite-radius,
positive-Hessian compactification saddle.  Its minimum has
$\mathcal K_{\rm int}=10.16\,\ell_P^{-4}$, below the adopted hard
quantum-boundary value $12\,\ell_P^{-4}$ of Sec.~\ref{sec:closure},
whereas the $(4,3)$ minimum lies beyond that boundary.  The alternative
saturated three-family sector $(1,5)$ is also locally stable; at these
fiducial couplings its well depth is $U_{\min}=-0.0384$, compared with
$U_{\min}=-0.0471$ for $(3,3)$.

The result also clarifies the role of torsion.  The torsion term alone,
being monotonic and give by $-1/(a_1^2a_2^4)$, would favor collapse.  Stability
arises from competition with quantized flux, whose steeper-inverse
powers dominate at sufficiently small radii.  Thus, the toy model
supports a nonsupersymmetric stabilization mechanism.  Finite Casimir,
nonzero Kaluza--Klein, and other corrections inside the admissible
domain are not computed here; however, Sec.~\ref{sec:closure} explains
why the Quantum Boundary Framework does not require an uncontrolled
continuation of the local derivative expansion to sub-cell scales.  The
short-distance prescription and the flux-dependent Kaluza--Klein tower
identified in Sec.~\ref{sec:bq-eval} remain the leading quantitative
uncertainties of the present truncation.

\section{Consequences for the nucleation weight}
\label{sec:nucleation}

Local stability does not determine which compactification sector carries
the larger quantum-cosmological weight.  In semiclassical quantum
cosmology, competing geometries are compared through saddle contributions
to the wave function or tunneling amplitude, so changes in the on-shell
energy or action can exponentially reorder otherwise admissible sectors
\cite{HartleHawking1983,Vilenkin1982}.  The derived vacuum term raises the stakes: an attractive
$N_{\rm fam}^2$ contribution rewards precisely the high-family sectors
that the flux cost is supposed to suppress.  We therefore ask whether
the three-family ordering of the companion synthesis survives once the
spin--torsion contribution is a computed coefficient rather than a
free phenomenological parameter, and if so, by how much.

In the thermal nucleation analysis of Ref.~\cite{ShayaUECKK2026}, the
sectors are ranked by their slice energies
$U_{\rm slice} = U_{\rm curv} + U_{\rm flux}
- \lambda_{\rm vac} N_{\rm fam}^2$.  With the vacuum term attractive,
three-family selection survives provided $\lambda_{\rm vac}$ lies
below the value at which the six-family sector $(2,5)$ overtakes
$(3,3)$:
\begin{equation}
\lambda_{\rm vac}^{\rm crit}
= \frac{U^{c+f}_{(2,5)} - U^{c+f}_{(3,3)}}{36-9}
= 0.461\,\Big(\frac{B}{16}\Big)\;M_P^4,
\label{eq:lam-crit}
\end{equation}
where $B\propto 1/g_{10}^2$ is the flux normalization.  Combining with
Eq.~\eqref{eq:lam-vac}, the viability of three-family dominance is a
\emph{joint condition on the ten-dimensional couplings}:
\begin{equation}
\boxed{\;
\frac{6}{\pi^4}\,\frac{\kappa_{10}}{V_6^{\rm slice}}
\;<\; 0.461\,\frac{B}{16}
\;\;\Longleftrightarrow\;\;
\kappa_4^{*} \;\lesssim\; 7.5\,\Big(\frac{B}{16}\Big)\,\ell_P^2,
\;}
\label{eq:viability}
\end{equation}
with $\kappa_4^{*}$ referred to the stabilized volume.  This is
satisfied in reduced-Planck conventions
($\kappa_4^{*} = 1$: $\lambda_{\rm vac} = 6/\pi^4 = 0.062$, margin
factor $7.5$) and violated in geometrized conventions
($\kappa_4^{*} = 8\pi$: $\lambda_{\rm vac} = 1.55$, excess factor
$3.4$).  Three-family dominance is therefore \emph{not} unconditional:
it is a sharp, falsifiable constraint on the gravitational
normalization of the underlying ten-dimensional theory, of exactly the
kind a fundamental completion must satisfy or fail.  At the viable
benchmark ($\kappa_4^{*}=1$, $B=16$) the nucleation ordering of
Ref.~\cite{ShayaUECKK2026} is preserved with $(3,3)$ dominating
$(2,5)$ by $e^{+68}$ and $(1,5)$ by $e^{+70}$; the $\mathcal{O}(1)$
uncertainties from $C_{\mathcal{B}}$ and the Kaluza--Klein tower
(Sec.~\ref{sec:bq-eval}) translate into an $\mathcal{O}(1)$ width of
the bound in Eq.~\eqref{eq:viability}.

\section{Compton--Schwarzschild closure and the quantum boundary}
\label{sec:closure}

The stability test above is nonsupersymmetric, so its most obvious
challenge is radiative stability near the Planck scale.  The conventional
effective-action expectation is a tower of higher-curvature operators;
indeed, higher-derivative gravity has long been studied precisely because
such terms alter the ultraviolet behavior \cite{Stelle1977}.  By
contrast, supersymmetric Calabi--Yau compactifications use special
holonomy and supersymmetry to control quantum corrections
\cite{Candelas1985}.  We therefore ask whether the geometric endpoint of
the Quantum Boundary Framework can play a different protective role by
removing the sub-cell configurations on which an indefinitely continued
derivative expansion would rely.

The \emph{Quantum Boundary Framework} of Ref.~\cite{ShayaUECKK2026}
provides an independent physical calibration of the ultraviolet end of
the geometric configuration space.  Its central assumption is not
that a continuum effective field theory remains valid to arbitrarily
short proper distance.  Rather, once the quantum uncertainty of a
geometric cell is of order the metric carried by that cell, the smooth
metric ceases to be an admissible path-integral variable.  Geometries
requiring structure on still smaller length scales are therefore not
assigned additional local corrections within the same continuum
manifold; they lie outside the supported configuration space.  In this
sense, the boundary removes the trans-boundary continuation of the
usual derivative expansion rather than regulating it by an arbitrarily
large momentum cutoff.

A complementary closure estimate follows from requiring that quantum
fluctuations become gravitationally self-confined.  For a homogeneous
FLRW interior with density $\rho$ and equation of state $w$, the
Kretschmann scalar is
\begin{equation}
\mathcal{K}
= \frac{64\pi^2 G^2\rho^2}{3}\,\Big[\,4 + (1+3w)^2\,\Big],
\label{eq:K-FLRW}
\end{equation}
minimized at $w = -1/3$ (where $\ddot a = 0$):
$\mathcal{K}_{\min} = 256\pi^2G^2\rho^2/3$.  The closure density, the
density at which the Hubble radius of the homogeneous FLRW interior
equals the cell scale $\ell_P$, is
\begin{equation}
\rho_{\rm close} = \frac{3c^2}{8\pi G\,\ell_P^{2}}\,,
\label{eq:rho-close}
\end{equation}
and substituting,
\begin{equation}
\mathcal{K}_c
= \frac{256\pi^2}{3}\,\frac{G^2}{c^4}\,\rho_{\rm close}^2
= \frac{256\pi^2}{3}\Big(\frac{3}{8\pi}\Big)^2\frac{1}{\ell_P^4}
= \frac{12}{\ell_P^4}\,.
\label{eq:Kc}
\end{equation}
This coincides with the Sobolev-failure threshold of the Quantum
Boundary Framework~\cite{ShayaUECKK2026}: the curvature at which the
system's own quantum fluctuations are gravitationally confined is the
curvature at which the metric ceases to be an admissible smooth
path-integral variable.  Taking the Compton scale at
$\sqrt2\,\ell_P$ ($\varpi = \sqrt{2}$ instead of $\varpi = 1$ in companion paper \cite{ShayaUECKK2026})
gives $\mathcal{K}_c = 3/\ell_P^4$; the
Sobolev value corresponds to the unreduced closure convention.

This interpretation is relevant to the usual concern that a
nonsupersymmetric compactification should receive an uncontrolled tower
of higher-derivative corrections.  A local operator containing
additional derivatives probes correspondingly shorter proper scales.
In an ordinary Wilsonian extrapolation one writes, schematically,
\begin{align}
\Gamma_{\rm loc}
= \int d^{10}x\sqrt{-g}\,\Big[&
\frac{R}{2\kappa_{10}}+c_2R^2+c_3\ell_P^2R^3
\nonumber\\
&+d_2\ell_P^2R\Box R+\cdots\Big].
\label{eq:derivative-tower}
\end{align}
and worries that when $\ell_P^2 R=\mathcal O(1)$ every term can become
comparable.  The Quantum Boundary Framework changes the premise of
that extrapolation.  Terms whose physical interpretation requires
resolving geometric variation on scales below the minimum supported
cell do not define additional sub-cell structure of the manifold,
because those metric configurations have zero support.  Moreover, the
higher the derivative order, the greater the Sobolev regularity needed
to interpret the corresponding local geometric functional.  If the
metric has already reached the regularity threshold at which the
Einstein--Hilbert configuration space terminates, operators demanding
still finer differentiability are not expected to remain independent
local degrees of freedom at the boundary.

We emphasize that this is a conjectural consequence of the Quantum
Boundary Framework, not a theorem that every finite higher-curvature
coefficient vanishes in the interior.  At radii comfortably above the
boundary, ordinary loop and matching corrections can still renormalize
the radion potential.  The stronger proposal is that the usual
argument for an \emph{uncontrolled trans-Planckian derivative tower at
nucleation} is inapplicable: the Feynman sum has no support on the
sub-cell geometries required to continue that tower to ever shorter
scales.  Since the $(3,3)$ saddle of Sec.~\ref{sec:stability} already
has a positive radion Hessian in the curvature--flux--torsion
truncation and lies on the admissible side of the boundary, we
speculate that its metastability can persist without the
boson--fermion cancellations normally supplied by unbroken
supersymmetry.  In this sense the Quantum Boundary Framework may provide
an alternative protection mechanism for the compactification, distinct
from the supersymmetric protection often invoked in Calabi--Yau model
building.  Establishing this claim quantitatively will require a
boundary-compatible calculation of the finite corrections that remain
inside the supported geometric domain.

\section{Discussion}
\label{sec:discussion}

The purpose of the preceding sequence was to separate four logically
distinct questions: the sign of the Lorentzian torsion interaction, the
existence of vacuum binding, the survival of that binding at finite
occupation, and the existence of a locally stable compactification
saddle.  Keeping those questions separate is essential because only the
short-distance evaluation invokes the Quantum Boundary Framework; the
channel algebra and separated-point correlator do not.  We now collect
what is unconditional, what is conditional, and what remains speculative.

The unconditional results of this paper are the separated-point
correlator Eq.~\eqref{eq:O-separated} and the channel factor
$R_{\rm ch} = -2$: the vacuum spin-torsion energy arises from quantum
current fluctuations, is dominated by the mixed
(spacetime$\times$internal) channels activated by the maximal spin
texture of the zero-mode subspace, and is attractive at every
supported separation.  The conditional results, the coefficient
$\lambda_{\rm vac}$ and the viability inequality
Eq.~\eqref{eq:viability}, follow from adopting the quantum-boundary
evaluation at $r = \ell_P$, and inherit $\mathcal{O}(1)$ uncertainties
from the smearing prescription and the Kaluza--Klein tower.

The physical picture that emerges is economical: one operator, one
sign, two epochs.  The vacuum piece supplies an attractive moduli term
at nucleation; in the coupled toy potential of Sec.~\ref{sec:stability}
quantized flux arrests the resulting contraction and produces a
positive-Hessian finite-radius minimum.  The occupation piece deepens
the binding after the cascade, while the single-particle self-energy
plays no role, the binding residing in the connected exchange
correlator.

The broader importance of these results becomes clearer when they are
placed beside the zero-mode theorem.  For the fixed product geometry
$S^2\times\mathbb{C}P^2$, the flux integers label a disjoint set of
compactification sectors and the index theorem converts those integers
into a family count.  This paper supplies the next map,
\begin{equation}
(m,n)\longrightarrow N_{\rm fam}
\longrightarrow -\lambda_{\rm vac}N_{\rm fam}^2,
\end{equation}
so the topological multiplicity of chiral matter feeds directly into
the radion dynamics.  In the $n=3$ sequence used in the UECKK analysis
\cite{ShayaUECKK2026}, the $S^2$ monopole support grows with $m^2$ at
the same time that the torsion attraction grows as
$N_{\rm fam}^2=m^2$, whereas the $\mathbb{C}P^2$ flux support remains
fixed.  The resulting imbalance explains the otherwise non-obvious
pattern in the two-modulus minima: increasing family number only
modestly changes the $S^2$ radius but progressively compresses the
$\mathbb{C}P^2$ factor.  At the fiducial couplings the $(3,3)$ sector
is the deepest locally stable minimum still inside the adopted
quantum-curvature boundary, while the $(4,3)$ minimum is pushed beyond
it.  The alternate three-family branch $(1,5)$ remains locally stable
but is less deeply bound in the same truncation.  These statements are
restricted to the explicit potential studied here; they are not a
classification of all compact manifolds or all flux sectors.

Ref.~\cite{ShayaUECKK2026} uses these quantitative results at the next
level of the argument.  The derived vacuum coefficient fixes the
family-dependent slice energy entering the proposed nucleation weight,
and Eq.~\eqref{eq:viability} states the coupling range in which the
three-family sector remains favored in that restricted ensemble.  The
same stability calculation shows why the family index is dynamically
relevant rather than merely numerological: on this class of manifolds,
the integer that counts chiral zero modes also controls the strength
of a gravitational vacuum interaction, thereby reshaping the
compactification potential.  That linkage; topology to chirality,
chirality to torsion energy, and torsion energy to moduli dynamics, is
the principal contribution of the present paper to the UECKK
synthesis.  The stronger claims about global ensemble selection and
the observed universe belong to Ref.~\cite{ShayaUECKK2026}; the
field-theoretic and two-modulus results on which those claims depend
are established here.

\bibliographystyle{apsrev4-2}
\bibliography{ueckk}

\end{document}